\documentstyle[11pt,appb,epsfig]{article}
\def\ra{\rightarrow}
\def\mrm{\rm}

\def\qqbar{\relax\ifmmode{\mrm{q}\overline{\mrm{q}}}%
                   \else${\mrm{q}\overline{\mrm{q}}}$\fi}
\def\bbbar{\relax\ifmmode{\mrm{b}\overline{\mrm{b}}}%
          \else${\mrm{b}\overline{\mrm{b}}}$\fi}
\def\nnbar{\relax\ifmmode{\nu\overline{\nu}}%
          \else${\nu\overline{\nu}}$\fi}
\def\llbar{\relax\ifmmode{\ell\overline{\ell}}%
          \else${\ell\overline{\ell}}$\fi}
\def\qqtn {\relax\ifmmode{\mrm{q\overline{q}}\tau\nu}%
                   \else${\mrm{q\overline{q}}\tau\nu}$\fi}
\def\qqln {\relax\ifmmode{\mrm{q\overline{q}}\ell\nu}%
                   \else${\mrm{q\overline{q}}\ell\nu}$\fi}
\def\qqen {\relax\ifmmode{\mrm{q\overline{q}e}\nu}%
                   \else${\mrm{q\overline{q}e}\nu}$\fi}
\def\qqmn {\relax\ifmmode{\mrm{q\overline{q}}\mu\nu}%
                   \else${\mrm{q\overline{q}}\mu\nu}$\fi}
\def\lnln {\relax\ifmmode{\ell\nu\ell\nu}%
                   \else${\ell\nu\ell\nu}$\fi}
\def\qqqq {\relax\ifmmode{\mrm{q\overline{q}q\overline{q}}}%
                   \else${\mrm{q\overline{q}q\overline{q}}}$\fi}
\newcommand{\epem}{\mbox{$\mrm{e^+e^-}$}}
\newcommand{\mpmm}{\mbox{$\mu^+\mu^-$}}
\newcommand{\tptm}{\mbox{$\tau^+\tau^-$}}

\newcommand{\Zz}{\mbox{${\mrm{Z}^0}$}}
\newcommand{\WW}{\mbox{$\mrm{W^+W^-}$}}
\newcommand{\wpwm}{\mbox{$\mrm{W^+W^-}$}}

\newcommand{\qq}{\mbox{$\mrm{q\overline{q}}$}}

\newcommand{\Mw}{\mbox{$\mrm{M}_{\mrm{W}}$}}
\newcommand{\roots}{\mbox{$\protect\sqrt{s}$}}
\newcommand{\rootsprime}{\mbox{$\protect\sqrt{s^\prime}$}}
\newcommand{\sprime}{\mbox{$\protect\sqrt{s^\prime}$}}

\newcommand{\Pmiss}{\not\!\mrm{P}}

\newcommand {\absctem} {\mbox{$|\cos\theta_{e^{-}} |$}}

\newcommand {\absctepm} {\mbox{$|\cos\theta_{e^{\pm}} |$}}

\newcommand{\Zo}{\mbox{$\mrm{Z}^{0}$}}

\newcommand{\Ho}{\mbox{$\mrm{H}^{0}$}}
\newcommand{\ee}{\mbox{$\mrm{e}^{+}\mrm{e}^{-}$}}
\newcommand{\mm}{\mbox{$\mu^{+}\mu^{-}$}}
\newcommand{\nn}{\mbox{$\nu \overline{\nu}$}}

\newcommand{\bb}{\mbox{$\mrm{b} \overline{\mrm{b}}$}}
\newcommand{\cc}{\mbox{$\mrm{c} \overline{\mrm{c}}$}}

\newcommand{\chpm}{\mbox{$\tilde{\chi}_{1}^{\pm}$}}

\newcommand{\nt}{\mbox{$\tilde{\chi}^{0}$}}
\newcommand{\neut}{\mbox{$\tilde{\chi}_{1}^{0}$}}
\newcommand{\neuttwo}{\mbox{$\tilde{\chi}_{2}^{0}$}}
\newcommand{\neutthree}{\mbox{$\tilde{\chi}_{3}^{0}$}}

\begin{document}
\begin{titlepage}
\begin{flushright}
  CR303 \\
  \today
\end{flushright}
\vspace*{1.25in}

\begin{center}
  {\large \bf Recent LEP2 Results from OPAL } \\
  \vspace*{0.25in}
  D.\ Glenzinski \\
  {\small Enrico Fermi Institute, University of Chicago \\
    Chicago Illinois, U.\ S.\ A.\ 60637 } \\
  \vspace*{0.50in}

  {\small Presented at the Cracow Epiphany Conference \\ 4-6 January 1997} \\
  \vspace*{0.25in}

  \vspace*{2.0in}

\end{center}

\normalsize
\begin{abstract}
  In 1996, after another set of upgrades, LEP began running for the first time 
  at center-of-mass energies above the W-pair threshold.  This new energy 
  regime offers a wide array of physics topics including tests of the Standard
  Model at higher energy scales, search physics, and W physics.  We summarize 
  the recent results from OPAL using 9.9~pb~$^{-1}$ of data collected at 
  $\roots = 161.3$~GeV from June to August 1996.
\end{abstract}
\end{titlepage}

\section{ {\bf Introduction}}
\label{intro}

In October of 1995 the very fruitful LEP1 physics program was officially 
brought to an end.  Later that same year, LEP saw its first running at 
center-of-mass energies significantly above the $\mrm{Z}^{0}$-peak 
($\roots \geq 130$~GeV).
In 1996, after another set of upgrades, LEP began running at center-of-mass 
energies above the W-pair threshold ($\roots > 2\mrm{M_{W}}$).  This new energy
regime offers a wide array of physics topics including tests of the Standard 
Model (SM) at higher energy scales, search physics, and W physics.  We 
summarize the recent results from OPAL using 9.9~pb~$^{-1}$ of data collected 
at $\roots = 161.3$~GeV from June to August 1996.  The OPAL detector is 
described in detail in reference~\cite{opal_ref}.

\section{ {\bf QCD Physics at LEP2 }}
\label{qcd}

\begin{figure}[p]
\begin{center}
\begin{minipage}[t]{5in}
\epsfxsize = 5in
\epsffile{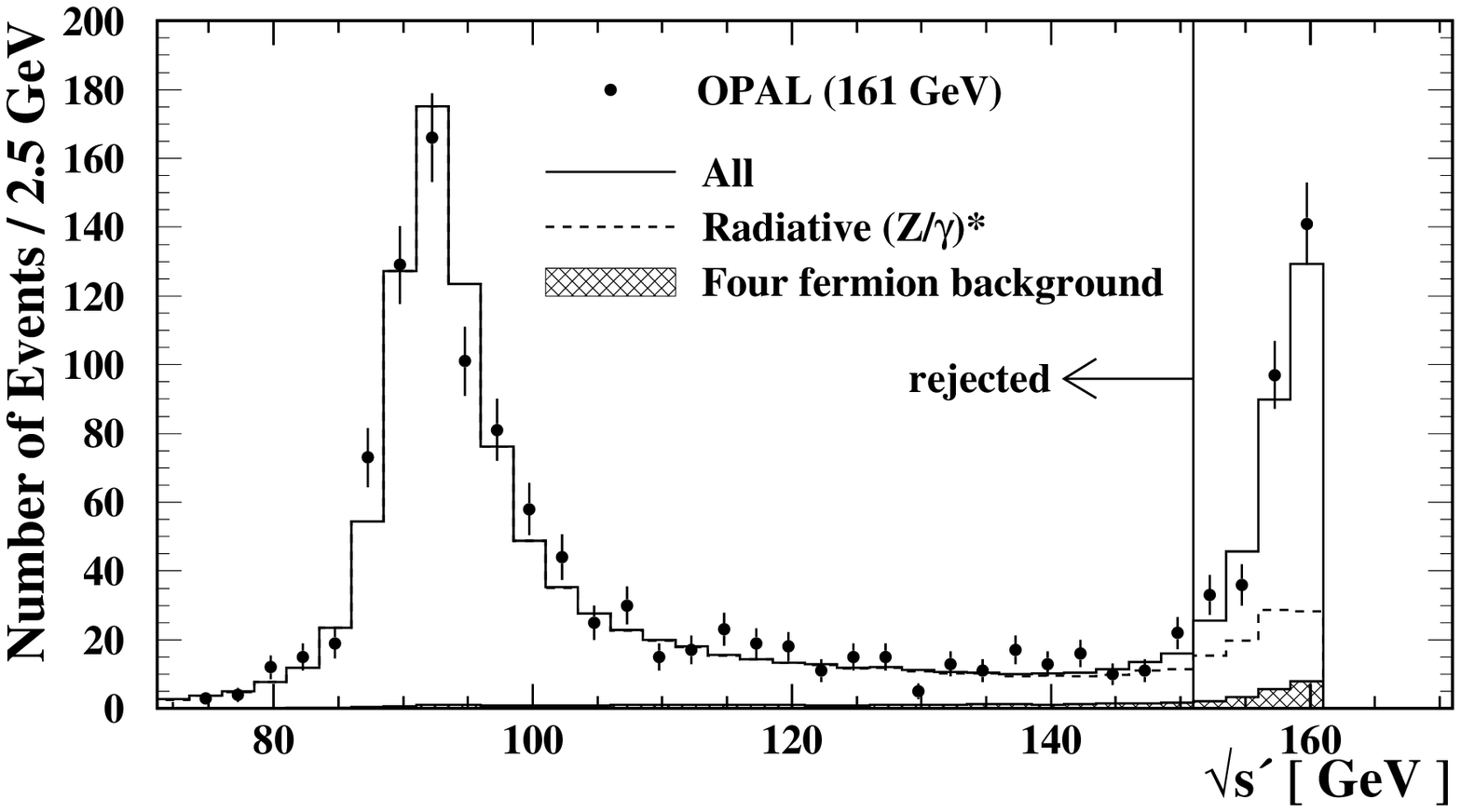}
\end{minipage}
\end{center}
\caption{Distribution of $\sprime$ for the data (full points) with statistical
errors.  The Monte Carlo predictions for the $\epem\ra \mrm{Z}/\gamma$ events 
(solid line), the radiative background (dashed line), and the  4-fermion 
background (hatched) are also shown.}
\label{qcd_sprime}
\end{figure}

\begin{figure}[p]
\begin{center}
\begin{minipage}[t]{5in}
\epsfxsize = 5in
\epsffile{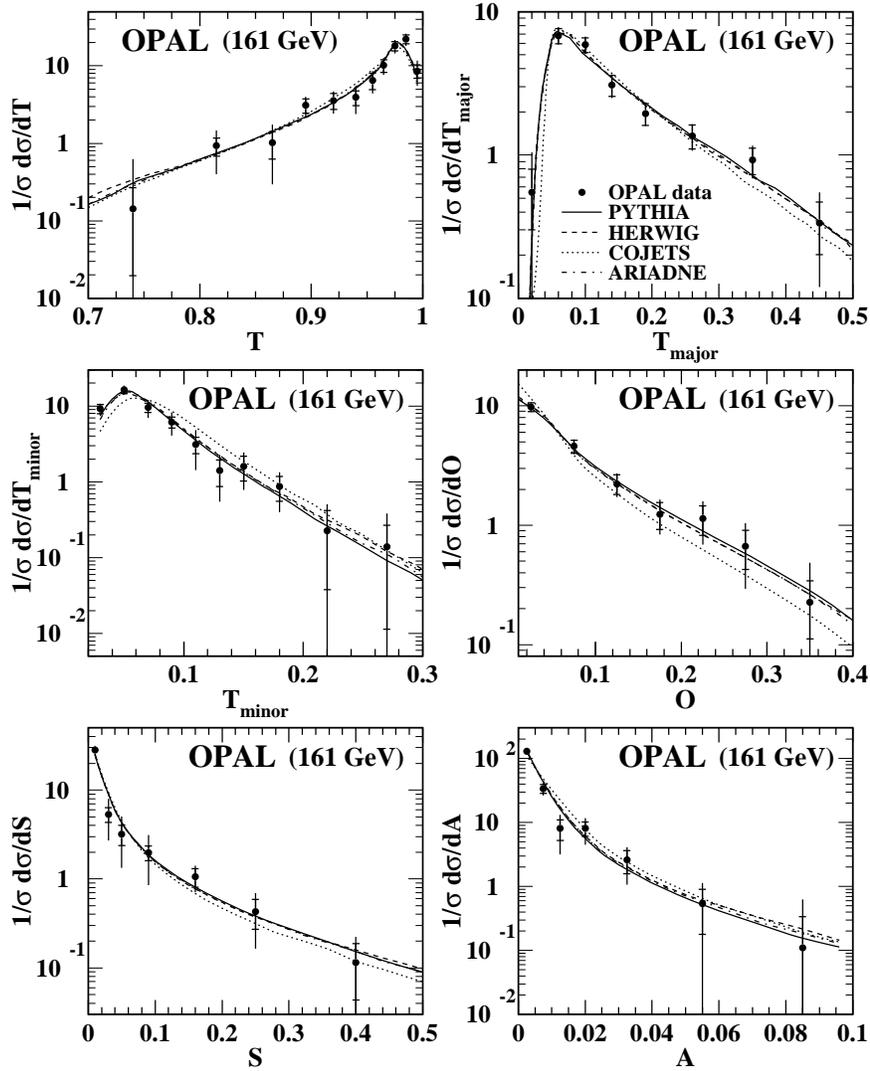}
\end{minipage}
\end{center}
\caption{Distributions of the event shape variables thrust ($T$), thrust major,
($T_{major}$), thrust minor ($T_{minor}$), oblateness ($O$), sphericity ($S$),
and aplanarity ($A$).  Experimental statisitical error bars are delimited by 
the small horizontal bars.  The total errors are shown by the vertical error 
lines.  Predictions from several Monte Carlo generators are also shown.}
\label{qcd_shapes}
\end{figure}

\begin{figure}[p]
\begin{center}
\begin{minipage}[t]{5in}
\epsfxsize = 5in
\epsffile{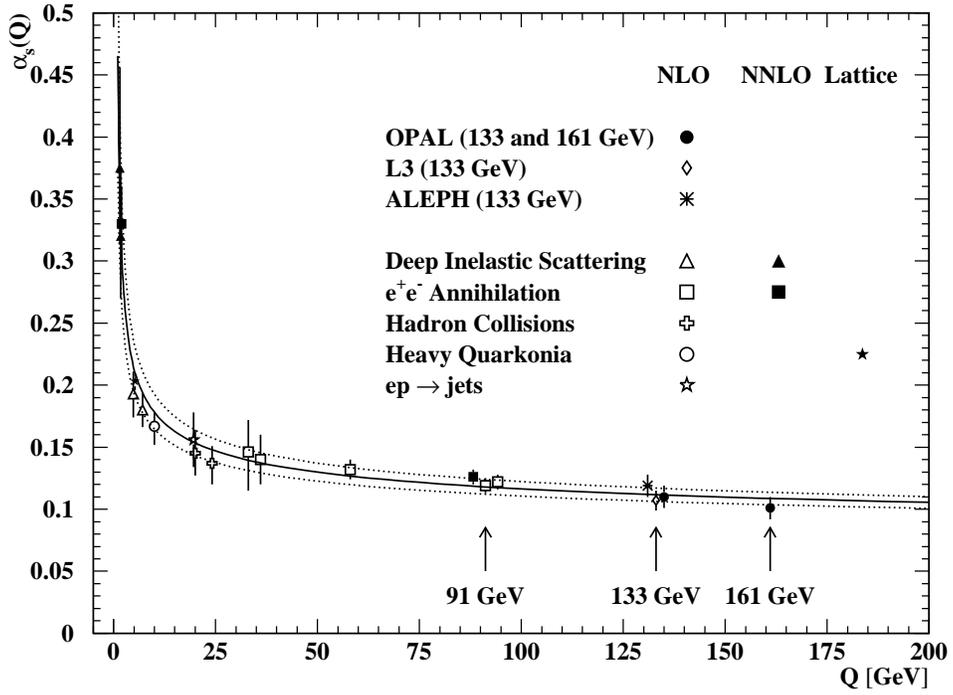}
\end{minipage}
\end{center}
\caption{Values of $\alpha_{\mrm{strong}}$ as a function of 
  energy~\protect\cite{as}.  The labels NLO and NNLO refer to the order of 
  calculation used.  NLO corresponds to ${\cal O}(\alpha_{\mrm{strong}}^{2})$ 
  in $\mrm{e^+e^-}$ annihilations, and NNLO to 
  ${\cal O}(\alpha_{\mrm{strong}}^{3})$.  The label Lattice refers to 
  $\alpha_{\mrm{strong}}$ values determined from lattice QCD calculations.  
  The curve shows the ${\cal O}(\alpha_{\mrm{strong}}^{3})$ QCD prediction for
  $\alpha_{\mrm{strong}}({\cal Q})$ using 
  $\alpha_{\mrm{strong}}(\mrm{M_{Z}})=0.118\pm 0.006$; the full line shows the 
  central value while the dotted lines indicate the variation given by the
  uncertainty.}
\label{qcd_alphas_running}
\end{figure}

\begin{figure}[p]
\begin{center}
\begin{minipage}[t]{5in}
\epsfxsize = 5in
\epsffile{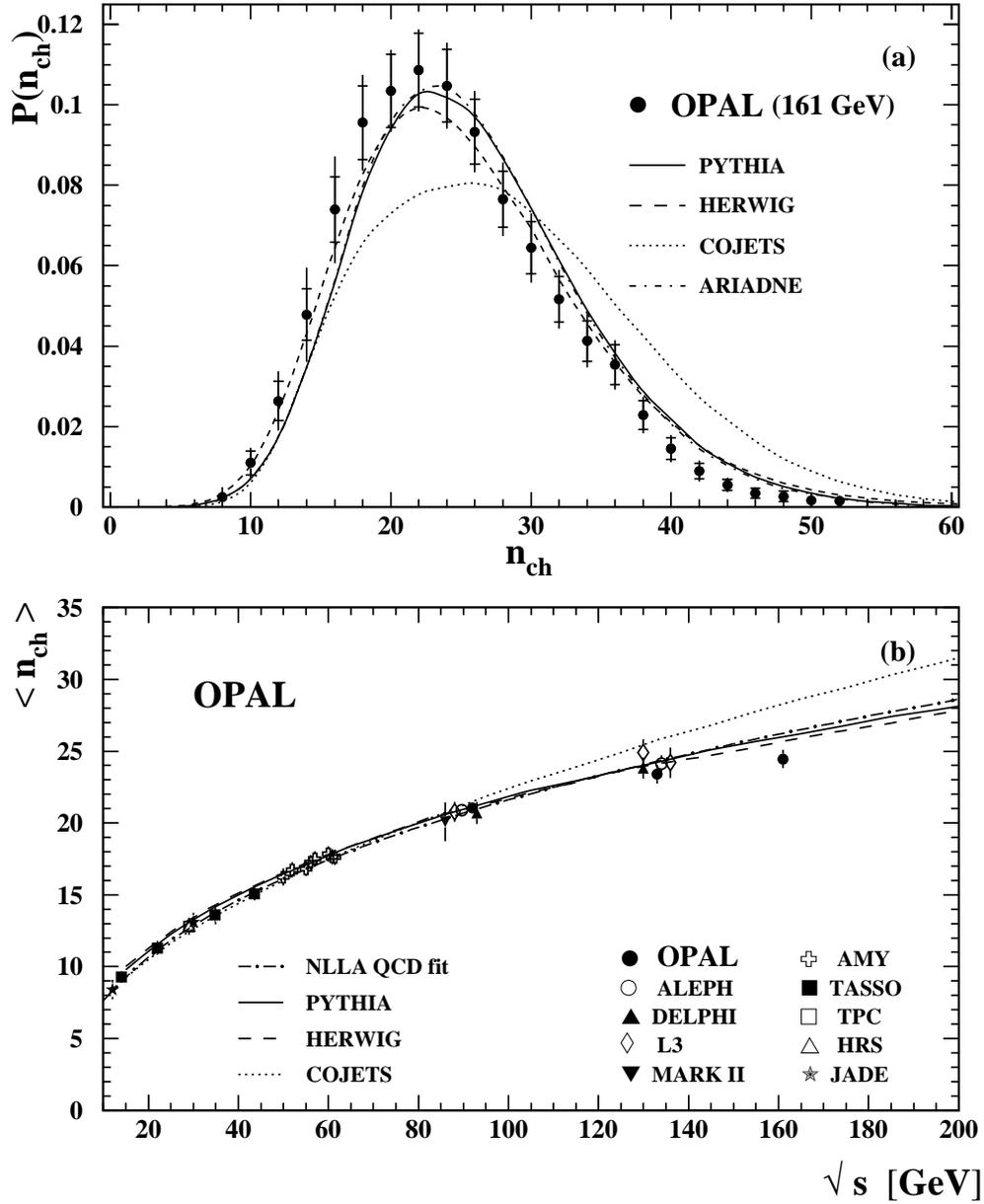}
\end{minipage}
\end{center}
\caption{(a) Corrected distribution of the charged particle multiplicity 
$n_{\mrm{ch}}$. Predictions from several Monte Carlo generators are also shown.
(b) Mean charged particle multiplicity measurements as a function of $\roots$.
The measurements are compared to a fit of the NLLA QCD prediction for the 
evolution of the charged particle multiplicity with $\roots$ and to the 
predictions of several Monte Carlo generators. }
\label{qcd_ave_nch}
\end{figure}

\begin{figure}[p]
\begin{center}
\begin{minipage}[t]{5in}
\epsfxsize = 5in
\epsffile{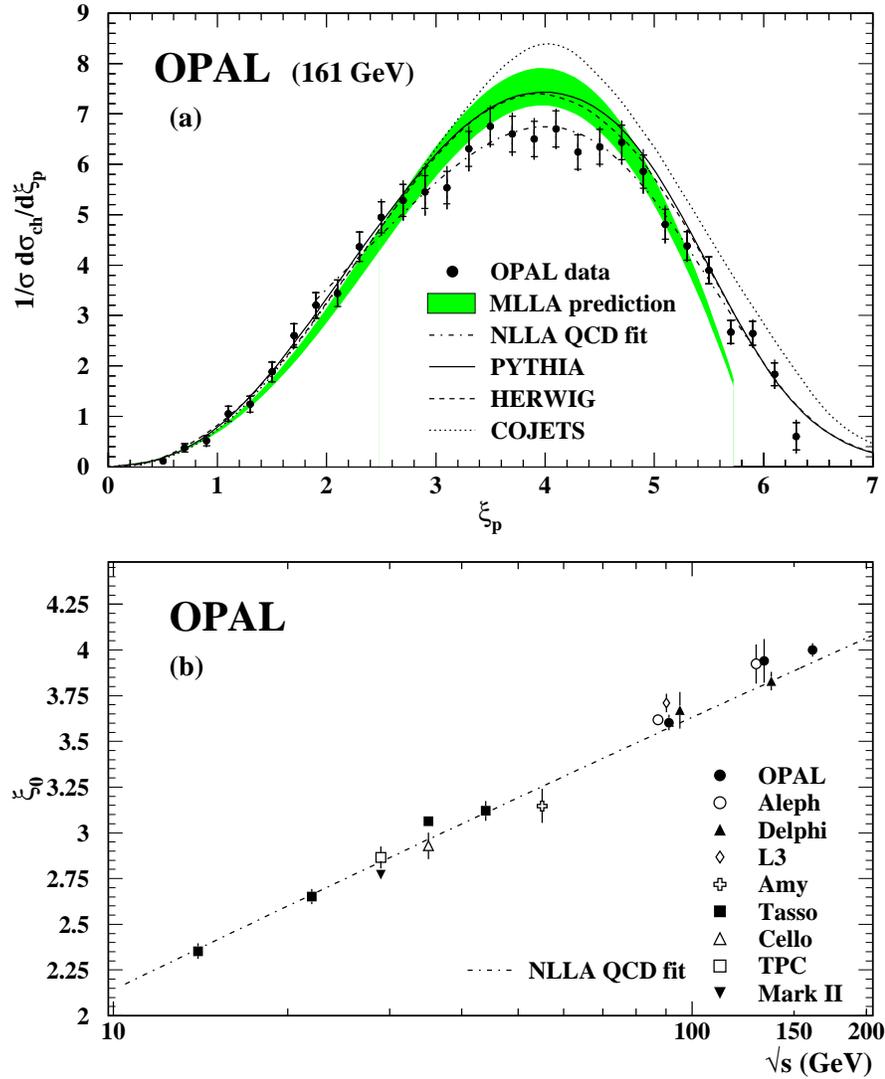}
\end{minipage}
\end{center}
\caption{(a) Distribution of $\xi_{p}=\ln (1/x_{p})$ for charged particles.  
Also shown are the MLLA QCD prediction and the predictions of several Monte
Carlo generators.  (b) Evolution of the position of the peak of the $\xi_{p}$
distribution, $\xi_{0}$, as a function of $\roots$, compared with a fit of a
NLLA QCD prediction up to and including the data points at $\roots = 130$~GeV.}
\label{qcd_xi_peak}
\end{figure}

The increased center-of-mass energy at LEP offers a new energy scale at which
to test the Standard Model, and in particular to test the predicted 
effects from the running of $\alpha_{\mrm{strong}}$ via QCD observables.  
At center-of-mass energies significantly above the $\mrm{Z}^{0}$-peak, 
initial state 
radiation (ISR) effects become large so that the effective center-of-mass of 
the $\epem$ interaction, $\rootsprime$, is less than the full energy 
available, $\roots = 2\mrm{E_{beam}}$.  In order to 
test the SM at a new energy scale it is necessary to differentiate full energy
events, in which $\rootsprime \approx \roots$, from events with a significant 
amount of initial state radiation - in particular the dominant 
``radiative return'' events in which $\rootsprime \approx \mrm{M_{Z^{0}}}$.
This is accomplished with a  kinematic fit that
calculates, for each event, an $\rootsprime$ using the visible energy and 
momentum measured in the event, and assuming energy and momentum conservation 
as described in reference~\cite{sprime_paper}.  The fit uses observed isolated 
photons and allows up to two unobserved ISR photons whose directions are taken
to be along the beam axis.  The resulting $\rootsprime$ distribution is shown 
in Figure~\ref{qcd_sprime} for events passing a high multiplicity 
pre-selection~\cite{tkmh_ref}. 
The so-called ``radiative return'' events give rise to a peak centered about 
the $\mrm{Z}^{0}$ mass.  To select full energy events we require that 
$\roots - \rootsprime < 10$~GeV.  This yields 307 events with an estimated 6\%
background from various 4-fermion processes (mostly $\wpwm$ events) and 
approximately a 5\% background from mismeasured radiative 
events~\cite{qcd_paper}.  For several QCD event shape variables we compare the
data to a variety of Monte Carlo generators, which employ a variety of
fragmentation schemes.  Figure~\ref{qcd_shapes} shows these comparisons
for the thrust, thrust major and minor, oblateness, sphericity, and aplanarity
event shape variables.  Using a combined fit
to a set of separate variables whose dependence on $\alpha_{\mrm{strong}}$ 
is predicted by NLLA QCD~\cite{qcd_theory_ref}, we measure
$\alpha_{\mrm{strong}}(161\mrm{GeV})=0.101 \pm 0.005(\mrm{stat.})
  \pm0.007(\mrm{syst.})$.  This measurement is compared to the 
QCD prediction~\cite{lattice_ref} in 
Figure~\ref{qcd_alphas_running}. In addition we measure the mean charged 
particle multiplicity to be 
$<\mrm{n_{ch}}>(161\mrm{GeV})=24.46\pm 0.45(\mrm{stat.})\pm 0.44(\mrm{syst.})$
and the position of the peak in the $\xi_{p} = \ln (1/x_{p})$ distribution 
to be 
$\xi_{0} (161\mrm{GeV}) = 4.00 \pm 0.03 (\mrm{stat.}) \pm 0.02 (\mrm{syst.})$.
These measurements are compared to data taken at lower center-of-mass energies
in Figures~\ref{qcd_ave_nch} and~\ref{qcd_xi_peak}.  These analyses are 
fully described in reference~\cite{qcd_paper}.

\section{ {\bf Two Fermion Physics at LEP2 }}
\label{2fermion}

\begin{figure}[p]
\begin{center}
\begin{minipage}[t]{5in}
\epsfxsize = 5in
\epsffile{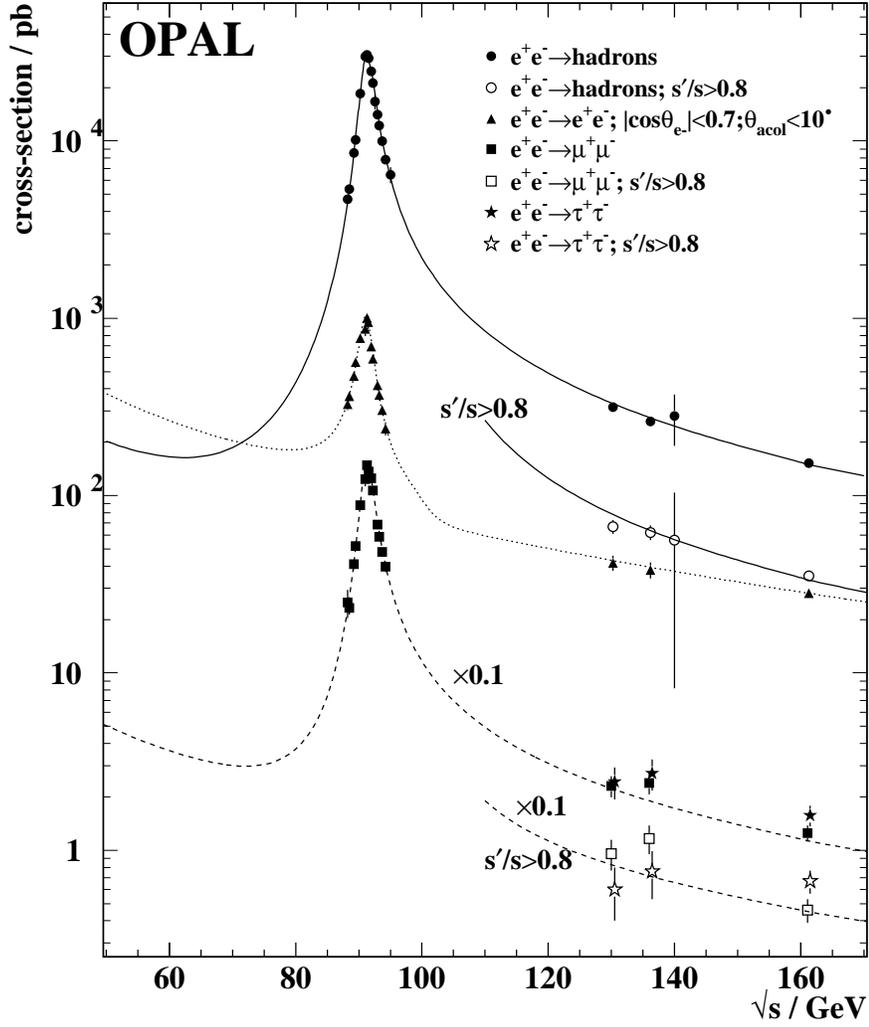}
\end{minipage}
\end{center}
\caption{Measured total cross-sections ($s^{\prime}/s > 0.01$) for different 
final states as a function of $\protect\sqrt{s}$.  The cross-sections for 
$\mu^+\mu^-$ and $\tau^+\tau^-$ production have been reduced by a factor of ten
for clarity. For the $\qqbar$, $\mu^+\mu^-$, and $\tau^+\tau^-$ final states, 
the cross-sections at high energies ($\protect\sqrt{s} \geq 130$~GeV) are also
shown for $s^{\prime}/s > 0.80$.  The curves show the SM predictions of 
ZFITTER and ALIBABA. }
\label{2ferm_xs_plot}
\end{figure}

\begin{figure}[p]
\begin{center}
\begin{minipage}[t]{5in}
\epsfxsize = 5in
\epsffile{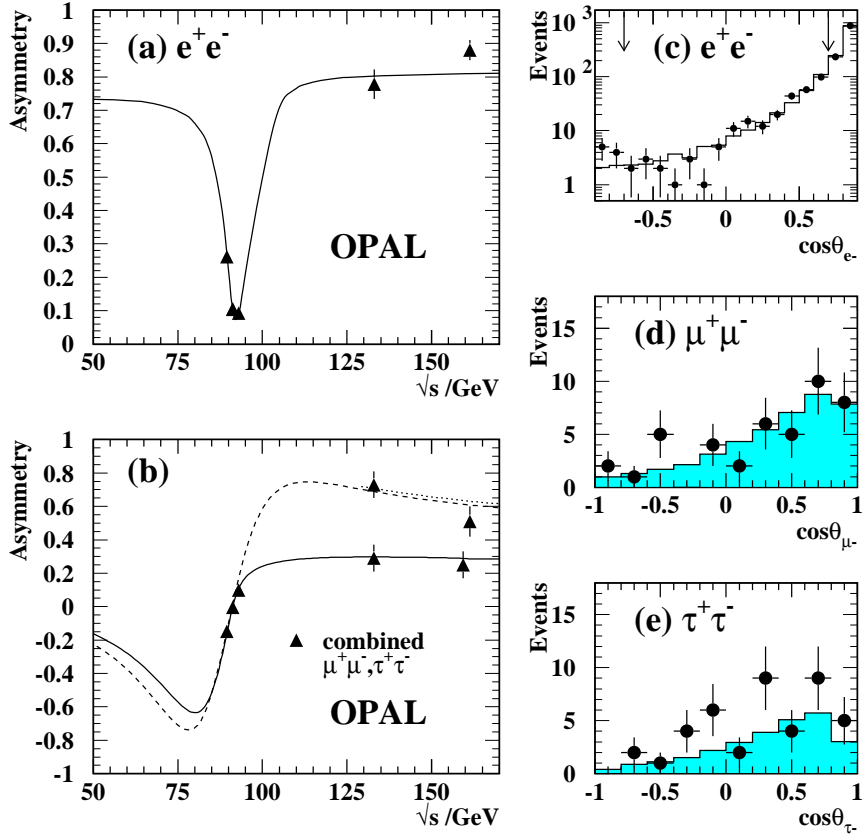}
\end{minipage}
\end{center}
\caption{(a) Measured forward-backward asymmetry for $\mrm{e^+e^-}$ pairs 
selected with $\left| \cos\theta_{e^-} \right| < 0.7$ and 
$\theta_{\mrm{acol}}<10^{\circ}$ as a function of $\protect\sqrt{s}$.  The 
curve shows the prediction of ALIBABA. (b) Measured asymmetries for the 
inclusive and exclusive samples of $\mu^+\mu^-$ and $\tau^+\tau^-$ combined.  
The curves show the predictions from ZFITTER for the inclusive (solid) and 
exclusive (dotted) selections as well as the Born-level expectation without QED
radiative effects (dashed).  The observed $\cos\theta$ distributions of the 
outgoing lepton are shown in (c) to (e) (points) and are compared with Monte 
Carlo expectations (histograms).  The arrows in (c) show the position of the 
cuts at $\left| \cos\theta_{e^-} \right| < 0.7$.}
\label{2ferm_afb}
\end{figure}

\begin{figure}[p]
\begin{center}
\begin{minipage}[t]{5in}
\epsfxsize = 5in
\epsffile{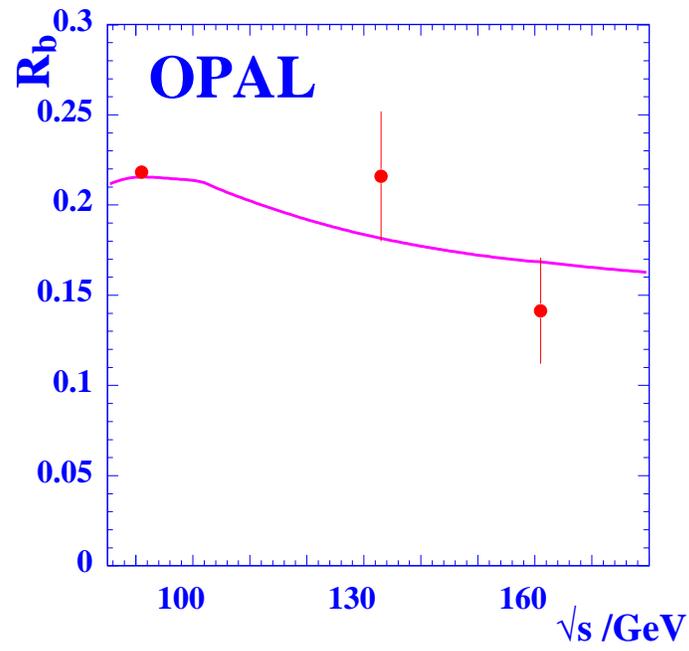}
\end{minipage}
\end{center}
\caption{$R_{b}$ as a function of $\protect\sqrt{s}$.  The points are the OPAL 
measurements and the solid line is the ZFITTER prediction.  The errors are
the quadrature sum of the statistical and systematic contributions.}
\label{2ferm_rb}
\end{figure}

Hadronic and leptonic two fermion events can also be used to test the SM at 
the higher energies available at LEP2.  We measure cross-sections and
asymmetries both including and excluding the dominant radiative return
events, $\epem\ra\gamma\mrm{Z^{0}}$, using the event selections described in 
Reference~\cite{2ferm_paper}. For the $\mpmm$, $\tptm$, and $\qqbar$ final 
states, we estimate the effective center-of-mass energy, $\sprime$, in a manner
similar to the one described above, in order to discriminate full energy 
events, $\sprime \approx \roots$, from the radiative return events, 
$\sprime \approx \mrm{M_{Z^{0}}}$.  An inclusive sample 
is defined by the cut $s^{\prime}/s > 0.01$ and an exclusive sample by the cut 
$s^{\prime}/s > 0.80$.  In the $\epem$ final state, due to the 
dominant $t$-channel production diagram, a definition of $\sprime$ as in the 
other final states is not meaningful.  Events with little radiation are 
therefore selected by a cut on the acollinearity angle between the electron and
positron, $\theta_{\mrm{acol}} = \pi - \theta_{e^{+}e^{-}}$.  A cut of
$\theta_{\mrm{acol}} < 10^{\circ}$ roughly corresponds to a cut of 
$s^{\prime}/s > 0.80$ for the $s$-channel contribution.  A sample with a
smaller $t$-channel contribution is identified by requiring the observed
electron to satisfy the condition $\left|\cos\theta_{e^{-}} \right| < 0.70$.
The observed number of events, measured cross-sections, and corresponding SM
predictions are shown in Table~\ref{2ferm_xs_table}.  These same results are
shown in comparison with lower energy data in Figure~\ref{2ferm_xs_plot}.
For the lepton-pair events we also measure the forward-backward asymmetry.
The results are shown graphically in Figure~\ref{2ferm_afb}, along with the 
lepton angular distributions, and agree with the SM predictions.  For the
non-radiative hadronic events we also measure $\mrm{R_{b}}$, the fraction of 
hadronic events
which decay into a $\mrm{b\overline{b}}$ pair, using a secondary-vertex tagging
method similar to the one described in Reference~\cite{opal_newest_rb}. We
find $\mrm{R_{b}}(161\mrm{GeV}) = 
        0.141 \pm 0.028(\mrm{stat.}) \pm 0.012(\mrm{syst.})$. 
Figure~\ref{2ferm_rb} plots $\mrm{R_{b}}$ as a function of center-of-mass 
energy for the three OPAL measurements at LEP1~\cite{opal_newest_rb}, LEP1.5, 
and this measurement.  All three measurements are within one standard deviation
of the SM expectation.

\begin{table}
\begin{center}
  \begin{tabular}{|l|c|c|c|} \hline \hline
    2 fermions & Sel.\ events & $\sigma$ (pb) & $\sigma^{\mrm{SM}}$ (pb) \\
    \hline
    Hadrons ($s'/s>0.01$)  & $1472$  & $152 \pm 4 \pm 2$        & $149$  \\
    Hadrons ($s'/s>0.8$)   & $370$   & $35.3 \pm 2.0 \pm 0.7$   & $33.2$ \\
    \hline
    \epem ($\absctem<0.70$, $\theta_{\mrm{acol}}<10^{\circ}$) 
                           & $285$   & $28.1 \pm 1.7 \pm 0.2$   & $28.1$ \\
    \epem ($\absctepm<0.96$, $\theta_{\mrm{acol}}<10^{\circ}$) 
                           & $4447$  & $435 \pm 7 \pm 6$        & $424$  \\
    \epem ($\absctepm<0.90$, $\theta_{\mrm{acol}}<170^{\circ}$) 
                           & $1582$  & $158 \pm 4 \pm 2$        & $153$  \\
    \hline
    \mpmm ($s'/s>0.01$)    & $98$    & $12.5 \pm 1.2 \pm 0.5$   & $11.3$ \\
    \mpmm ($s'/s>0.8$)     & $44$    & $4.6 \pm 0.7 \pm 0.2$    & $4.5$  \\
    \hline
    \tptm ($s'/s>0.01$)    & $64$    & $15.7 \pm 2.0 \pm 0.7$   & $11.3$ \\
    \tptm ($s'/s>0.8$)     & $43$    & $6.7 \pm 1.0 \pm 0.3$    & $4.5$  \\
    \hline \hline
  \end{tabular}
\end{center}
\caption{ Numbers of events and measured cross-sections at 
  $\protect\sqrt{s} = 161.3$~GeV.  For the cross-sections, the first error is 
  statistical and the second is systematic.  The last column shows the SM
  cross-section predictions from ZFITTER~\protect\cite{zfitter_ref} 
  ($\mrm{q}\overline{\mrm{q}}$, $\mu^+\mu^-$, and $\tau^+\tau^-$) and 
  ALIBABA~\protect\cite{alibaba_ref} $\left( \mrm{e^+e^-} \right)$. }
\label{2ferm_xs_table}
\end{table}

These analyses are described in detail in Reference~\cite{2ferm_paper}.

\section{ {\bf Search Physics at LEP2 }}
\label{searches}

The increased center-of-mass energy at LEP2 opens up an entirely new region of
parameter space for a variety of possible new physics signatures.  OPAL has a 
wide and varied program in order to be as sensitive to as many topologies as 
possible.  The principal signature is that of missing momentum ($\Pmiss$) plus
a pair of acoplanar jets, leptons, or some combination thereof. These simple 
topologies allow sensitivity to SM Higgs, Supersymmetric (SUSY) Higgs, 
chargino, neutralino, slepton, stop, sbottom, excited lepton, and both charged
and neutral heavy lepton  production processes.  By including 4-jet topologies,
and exploiting for particular search channels the presence of hard photons, 
b-jets, and/or resonances, OPAL achieves reasonable efficiencies over a large 
parameter space for many models.  No significant excess is observed in any of 
our searches, and a variety of limits are set at the $95\%$ confidence level.
Although there are OPAL results for all of the above mentioned 
processes~\cite{searches}, I will only discuss here the results obtained from 
the SM Higgs, the chargino and neutralino, and the anomalous 4-jet production
searches.

\subsection{\bf Search for the Standard Model Higgs Boson}
\label{smhiggs}

The higher center-of-mass energy available at LEP2 increases the sensitivity 
of the search for a SM Higgs boson.
At this centre-of-mass energy, the main production process for the SM Higgs 
boson is $\epem \rightarrow \mrm{Z^{0}H^{0}}$. The dominant decay is
$\Ho\ra\bb$, with a branching ratio of approximately 86\%.  Other relevant 
decay modes are: $\Ho\ra\tptm$ (8\%), $\Ho\ra\cc$ (4\%), and 
$\Ho\ra$ gluons (2\%) \cite{spira}. In the mass range of interest, these 
branching ratios exhibit only a mild dependence on the Higgs boson mass.

The OPAL search is sensitive to the principal final state topologies,
namely: (i) the four-jet channel, $\ee\ra\Zo\Ho\ra\qq\bb$; (ii) the missing 
energy channel, mainly from $\ee\ra\Zo\Ho\ra\nn\qq$, but including a small
contribution from the $\WW$ fusion process $\ee\ra\nn\Ho$; (iii) the tau 
channels, $\ee\ra\Zo\Ho\ra\tptm\qq$ and $\qq\tptm$; and (iv) the electron and 
muon channels, predominantly from  $\ee\ra\Zo\Ho\ra\ee\qq$ and $\mm\qq$, but 
including a small contribution from the $\Zz\Zz$ fusion process 
$\ee\ra\ee\Ho$. These topologies account for about 95\% of all Higgs boson 
final states.

\begin{figure}[p]
\begin{center}
\begin{minipage}[t]{5in}
\epsfxsize = 5in
\epsffile{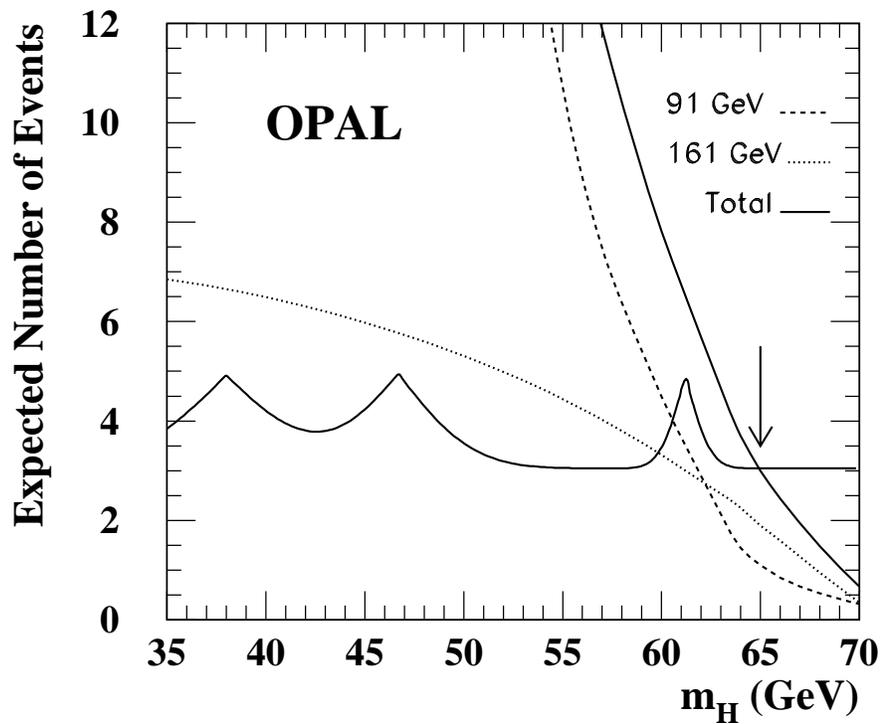}
\end{minipage}
\end{center}
\caption{Expected number of events as a function of the Higgs boson mass, 
$\mrm{M_{H}}$, for the search at $\roots = 161.3$~GeV (dotted), and at 
$\roots \approx \mrm{M_{Z}^{0}}$ (dashed).  Combining the searches yields the 
expectation given by the falling solid line. The 95\% confidence level upper 
limit in the presence of three candidate events is given as the solid 
horizontal curve.  The intersection of the two solid curves, indicated by the 
arrow, determines our 95\% confidence level lower limit on the Higgs boson 
mass.}
\label{smhigg_limit_plot}
\end{figure}

Table~\ref{smhiggs_table} lists the typical efficiency for each channel and
gives the observed and expected number of background events for the 
approximately $10$~pb$^{-1}$ of data collected at $\roots = 161$~GeV.  The 
observations are in good agreement with the number of expected events from 
Standard Model background processes. By combining this data with data taken
at $\roots \approx \mrm{M_{Z^{0}}}$, we derive a lower limit on the mass of the
Higgs boson of $\mrm{M_{H^{0}}} > 65.0$~GeV at the $95\%$ confidence level.  
This limit is shown in Figure~\ref{smhigg_limit_plot}.  
\begin{table}
\begin{center}
  \begin{tabular}{|ccccc|} \hline
    $\mrm{H}^{0}$  &$\mrm{Z}^{0}$   &$\mrm{Eff}\left(\%\right)$ 
                   &$\mrm{N_{bkg}}$ &$\mrm{N_{obs}}$           \\ \hline \hline
    $\bbbar$       &$\qqbar$        &$23$  &$0.8$    &$1$   \\ 
    $\qqbar$       &$\nnbar$        &$46$  &$0.9$    &$1$   \\ 
    $\qqbar$       &$\tptm $        &$20$  &$0.2$    &$0$   \\ 
    $\qqbar$       &$\llbar$        &$65$  &$0.1$    &$0$   \\ \hline
  \end{tabular}
\end{center}
\caption{ The channel by channel efficiency for a Higgs boson of mass 
  $\mrm{M_{H}}=65$~GeV is given along with the number of expected background 
  events from SM processes.  The number of observed events is also given and
  is consistent with the background expectation.}
\label{smhiggs_table}
\end{table}

A more detailed description of this analysis can be found in 
Reference~\cite{sm_higgs_paper}.

\subsection{\bf Search for Chargino and Neutralino Production}
\label{chargino_neut}

We perform a direct search for the pair production of charginos and 
neutralinos, whose existence is predicted in SUSY theories.  Charginos, 
$\tilde{\chi}^{\pm}_{j}$, are the mass eigenstates formed by the mixing of the 
fields of the fermionic partners of the charged gauge bosons (winos) and those
of the charged Higgs bosons (charged higgsinos).  Fermionic partners of the 
photon, the $\mrm{Z}$ boson, and the neutral Higgs bosons mix to form the mass 
eigenstates called neutralinos, $\tilde{\chi}^{0}_{i}$.  In each case, the 
index $j$ or $i$ increases with increasing mass.  

If charginos are light enough, they can be pair produced in \epem collisions 
through $\gamma$ or $\mrm{Z}^{*}$ exchange in the $s$-channel and sneutrino 
($\tilde{\nu}$) exchange in the $t$-channel.  Neutralino pairs 
($\tilde{\chi}^{0}_{i}\tilde{\chi}^{0}_{j}$) can be produced through an
$s$-channel $\gamma$ or $\mrm{Z}^{*}$ exchange, or by $t$-channel selectron
($\tilde{e}$) exchange.

We assume that the lightest neutralino, $\tilde{\chi}^{0}_{1}$, is the lighest
supersymmetric particle and that R-parity is conserved.  Experimentally these 
assumptions have the consequence that the $\tilde{\chi}^{0}_{1}$ is stable and
invisible. The lightest chargino, $\tilde{\chi}^{\pm}_{1}$, can then decay via
$\tilde{\chi}^{\pm}_{1} \ra \tilde{\chi}^{0}_{1} \ell^{+}\nu$ or 
$\tilde{\chi}^{\pm}_{1} \ra \tilde{\chi}^{0}_{1} \mrm{q\overline{q}^{\prime}}$
while the $\tilde{\chi}^{0}_{2}$ can then decay into the final states
$\tilde{\chi}^{0}_{1} \nu\overline{\nu}$, 
$\tilde{\chi}^{0}_{1} \ell^{+}\ell^{-}$, or
$\tilde{\chi}^{0}_{1} \mrm{q\overline{q}}$.
These assumptions  have the additional consequence that since events of the 
type $\epem \ra \tilde{\chi}^{0}_{1} \tilde{\chi}^{0}_{1} \gamma$ would suffer 
from a large irreducible background from the standard model process
$\epem \ra \nu\overline{\nu}\gamma$, we can only achieve a reasonable 
sensitivity for events of the type
$\epem \ra \tilde{\chi}^{0}_{2} \tilde{\chi}^{0}_{1}$, and 
$\epem \ra \tilde{\chi}^{0}_{3} \tilde{\chi}^{0}_{1}$.  Note that the final 
state kinematics, and therefore the detection efficiencies, will depend on the
mass difference between the chargino and the lightest neutralino,
$\Delta \mrm{M}_{\pm} = \mrm{M}(\chpm) - \mrm{M}(\neut)$. Similarly for the 
neutralinos, whose detection efficiency will depend upon the mass difference
$\Delta \mrm{M}_{0} = \mrm{M}(\neuttwo) - \mrm{M}(\neut)$. 

\begin{figure}[p]
\begin{center}
\mbox{\epsfig{file=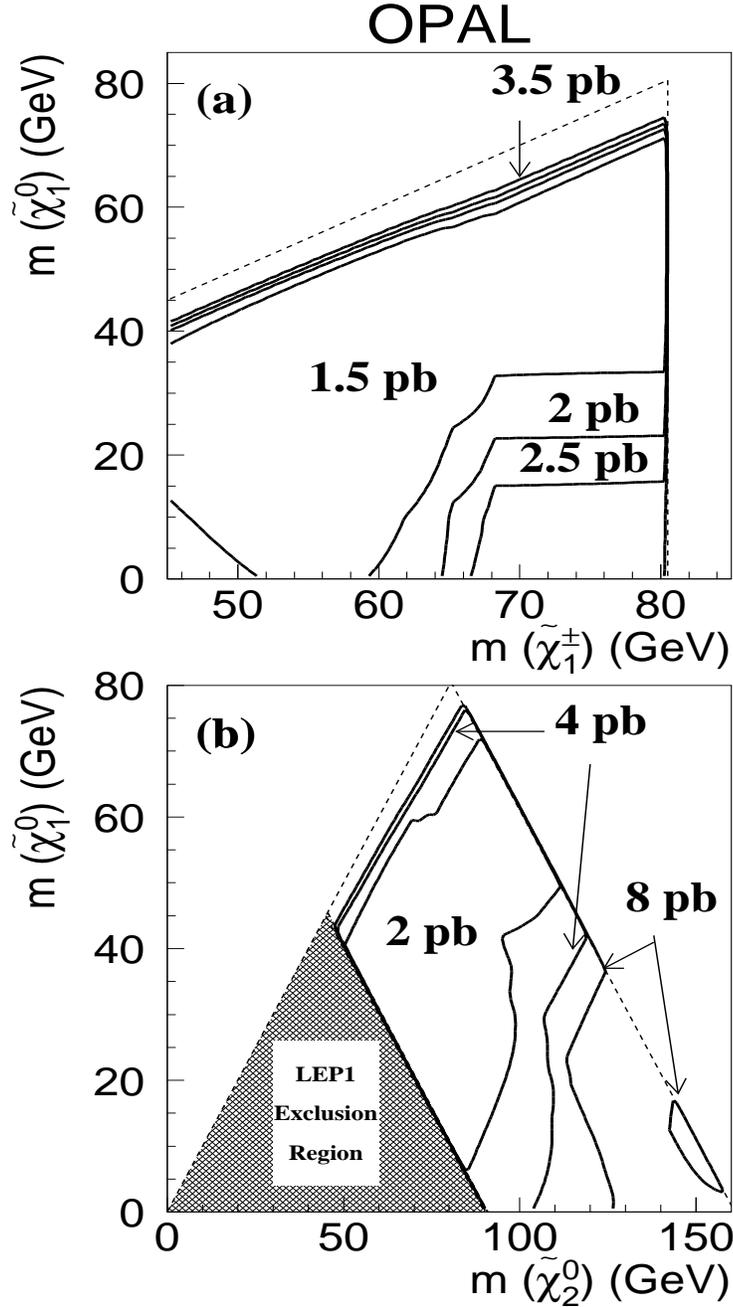,height=18.0cm,width=12.0cm}}
\caption{ The 95\% confidence level production cross-section contours for
(a) $\mrm{e^+e^-} \ra \tilde{\chi}_{1}^{+} \tilde{\chi}_{1}^{-}$, and 
(b) $\mrm{e^+e^-} \ra \tilde{\chi}_{1}^{0} \tilde{\chi}_{2}^{0}$ assuming the 
decays $\tilde{\chi}^{\pm}_{1} \ra \tilde{\chi}^{0}_{1}\mrm{W}^{*\pm}$ and
$\tilde{\chi}^{0}_{2} \ra \tilde{\chi}^{0}_{1}\mrm{Z}^{*}$ occur with
$100\%$ branching fraction.  These limits have been obtained by combining the
results of the $\roots = 161$~GeV and $\roots = 130-136$~GeV 
analyses~\protect\cite{ch_neut_lep15}.}
\label{charg_neut_xs_limit_plot}
\end{center}
\end{figure}

Typical efficiencies for the various final state topologies are given in 
Table~\ref{charg_neut_table} along with the total number of observed and
expected events.  No significant excess is observed.  
Table~\ref{charg_neut_limits} gives the $95\%$ confidence level lower limits
that we extract in the context of the Minimal Supersymmetric Standard Model 
(MSSM) theory assuming that $\Delta\mrm{M_{\pm}} > 10$~GeV and
 $\Delta\mrm{M_{0}} > 10$~GeV.  Figure~\ref{charg_neut_xs_limit_plot} shows
the $95\%$ confidence level upper limit cross-section contours for
$\tilde{\chi}^{+}_{1}\tilde{\chi}^{-}_{1}$ and 
$\tilde{\chi}^{0}_{2}\tilde{\chi}^{0}_{1}$ production assuming the decays
$\tilde{\chi}^{\pm}_{1} \ra \tilde{\chi}^{0}_{1}\mrm{W}^{*\pm}$ and
$\tilde{\chi}^{0}_{2} \ra \tilde{\chi}^{0}_{1}\mrm{Z}^{*}$ occur with
$100\%$ branching fraction.
\begin{table}
\begin{center}
  \begin{tabular}{|cc|c|c|c|} \hline
           & &\multicolumn{3}{l|}{$\Pmiss + \mrm{Acoplanar:}$}  \\
           & &$\mrm{jets}$ &$\mrm{jets}+\ell$ &$\mrm{leptons}$  \\ 
           \hline
           &small$\Delta\mrm{M}_{\pm}$  
                                  &$10-20\%$   &$15-20\%$   &$11-24\%$    \\
   $\chpm$ &large $\Delta\mrm{M}_{\pm}$ 
                                  &$40-60\%$   &$30-65\%$   & ---         \\
           &BR                    &$45\%$      &$45\%$      &$10\%$       \\ 
           \hline
           &small$\Delta\mrm{M}_{0}$  
                                  &$5-20\%$    & ---        &$5-22\%$     \\
   $\nt$   &large $\Delta\mrm{M}_{0}$ 
                                  &$20-45\%$   & ---        & ---         \\
           &BR                    &$40-80\%$   &            &$20-60\%$    \\ 
           \hline \hline
     \multicolumn{5}{|c|}{$0.7 \pm 0.2$ expected background} \\
     \multicolumn{5}{|c|}{$2$ events observed} \\
     \hline  
  \end{tabular}
\end{center}
\caption{ Typical efficiencies for the various topologies used in the 
  chargino and neutralino searches along with the number of expected events
  from SM processes.  The number of observed events is consistent with this
  background expectation.}
\label{charg_neut_table}
\end{table}

\begin{table}
\begin{center}
  \begin{tabular}{|c|c|c|c|c|} \hline
    Mass & \multicolumn{2}{c|}{$\tan \beta = 1.5$}
         & \multicolumn{2}{c|}{$\tan \beta = 35$} \\ 
         \cline{2-5}
    (GeV)&Min. $m_0$ &$m_0$=$1$~TeV 
         &Min. $m_0$ &$m_0$=$1$~TeV   \\ 
         \hline 
    $\mrm{M}_{\chpm}$ & $62.0$ & $78.5$ & $66.5$  &$78.8$   \\
    $\mrm{M}_{\neut}$ & $12.0$ & $30.3$ & $35.8$  &$41.3$   \\
    $\mrm{M}_{\neuttwo}$ & $45.3$ & $51.9$ & $67.2$  &$80.0$   \\
    $\mrm{M}_{\neutthree}$ & $86.3$ & $94.3$ & $112.5$ &$112.5$  \\
    \hline
  \end{tabular}
\end{center}
\caption{ Lower limits on chargino and neutralino masses at the $95\%$ 
  confidence 
  level in the context of the MSSM and assuming $\Delta M > 10$~GeV.  The
  limits are derived for both the minimal $m_0$ consistent with present 
  experimental constraints and $m_0 = 1$~TeV for the two cases 
  $\tan \beta = 1.5$ and $\tan \beta = 35$. The limits on the $\neutthree$ mass
  are obtained mainly from excluded regions in the MSSM parameter space 
  resulting from the direct search for lighter neutralinos and $\chpm$.}
\label{charg_neut_limits}
\end{table}

A more detailed description of this analysis can be found in 
Reference~\cite{charg_neut_paper}.

\subsection{ {\bf Anomalous Four-Jet Production }}
\label{fourjets}

\begin{figure}[phtb]
\begin{center}
\mbox{\epsfig{file=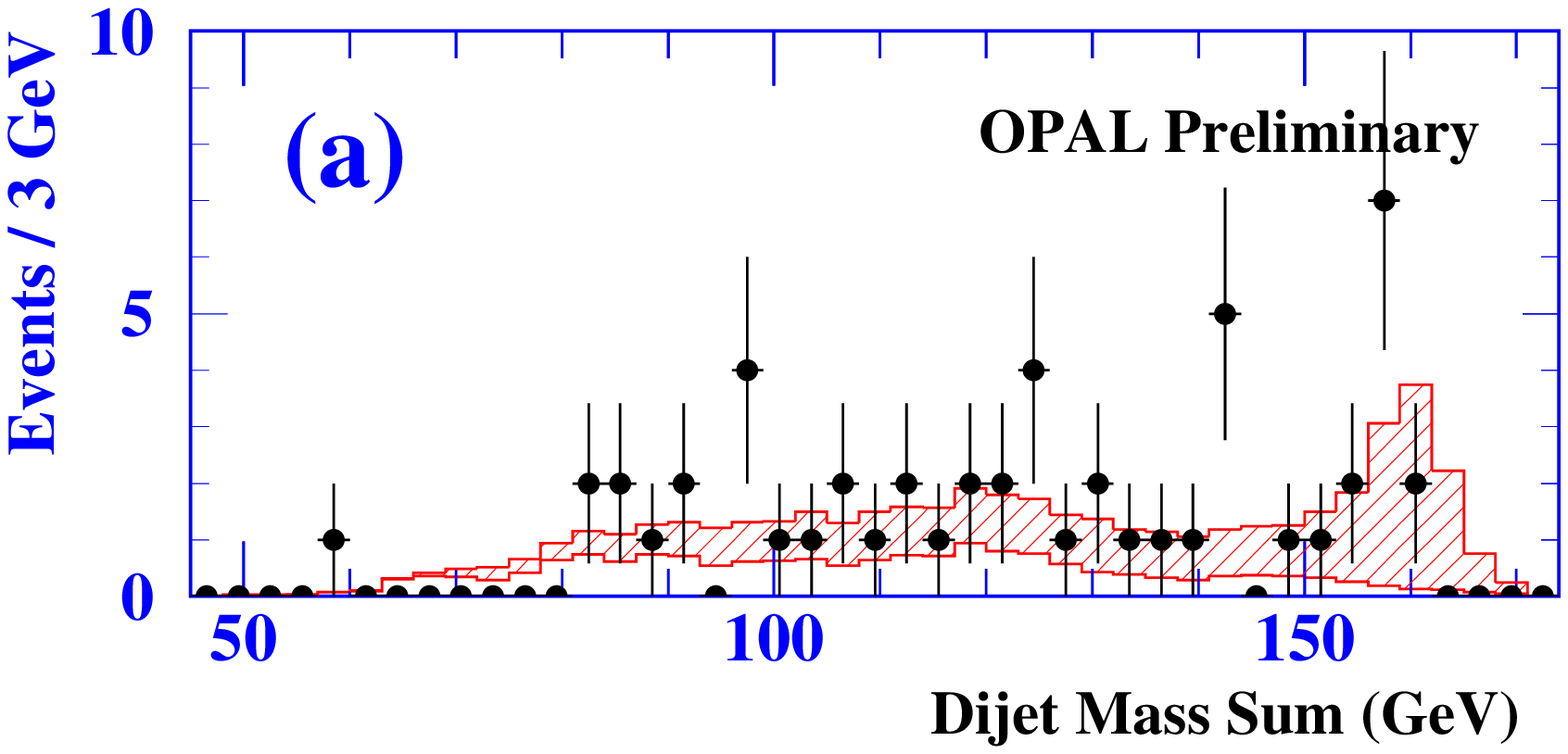,height=5cm,width=12.0cm}}
\mbox{\epsfig{file=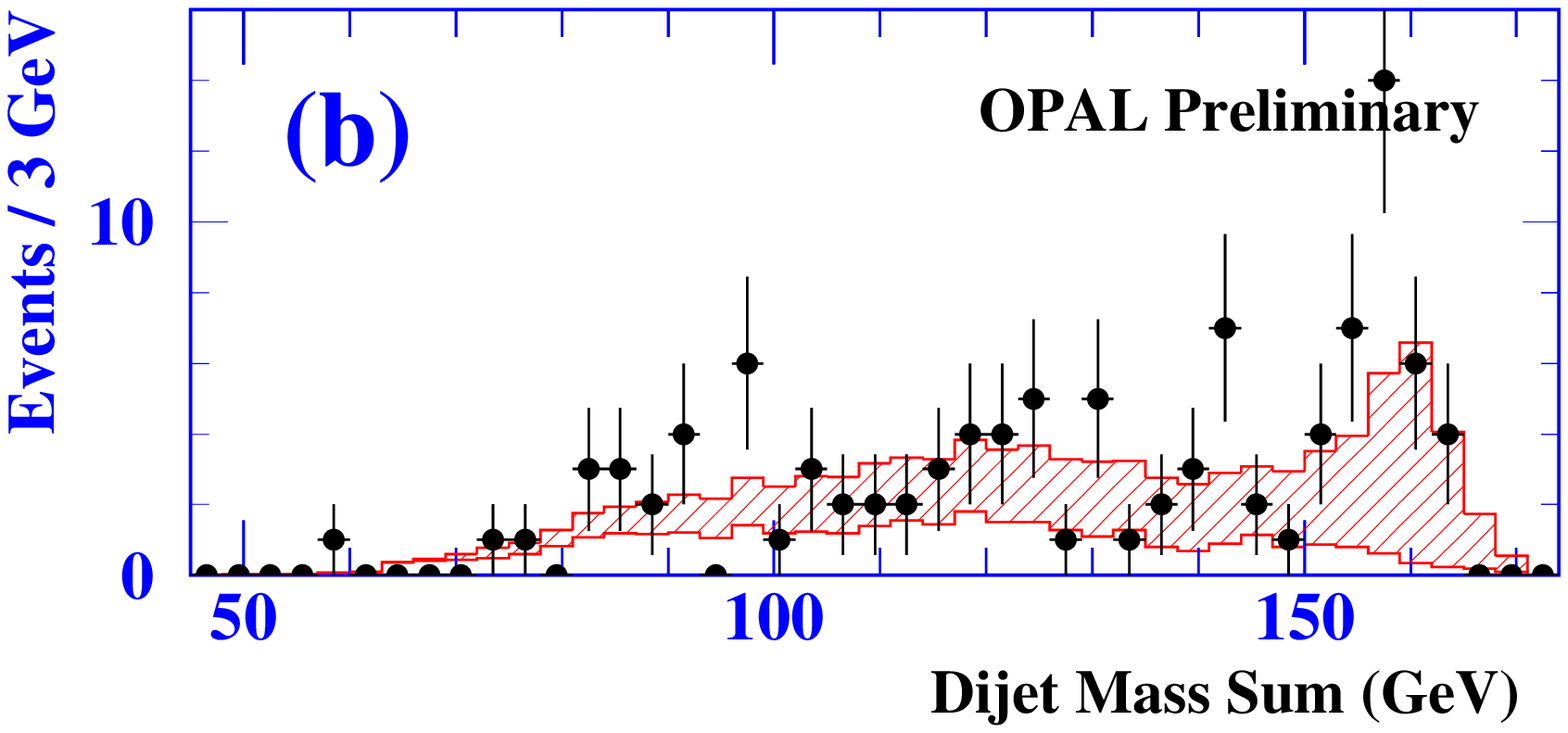,height=5cm,width=12.0cm}}
\mbox{\epsfig{file=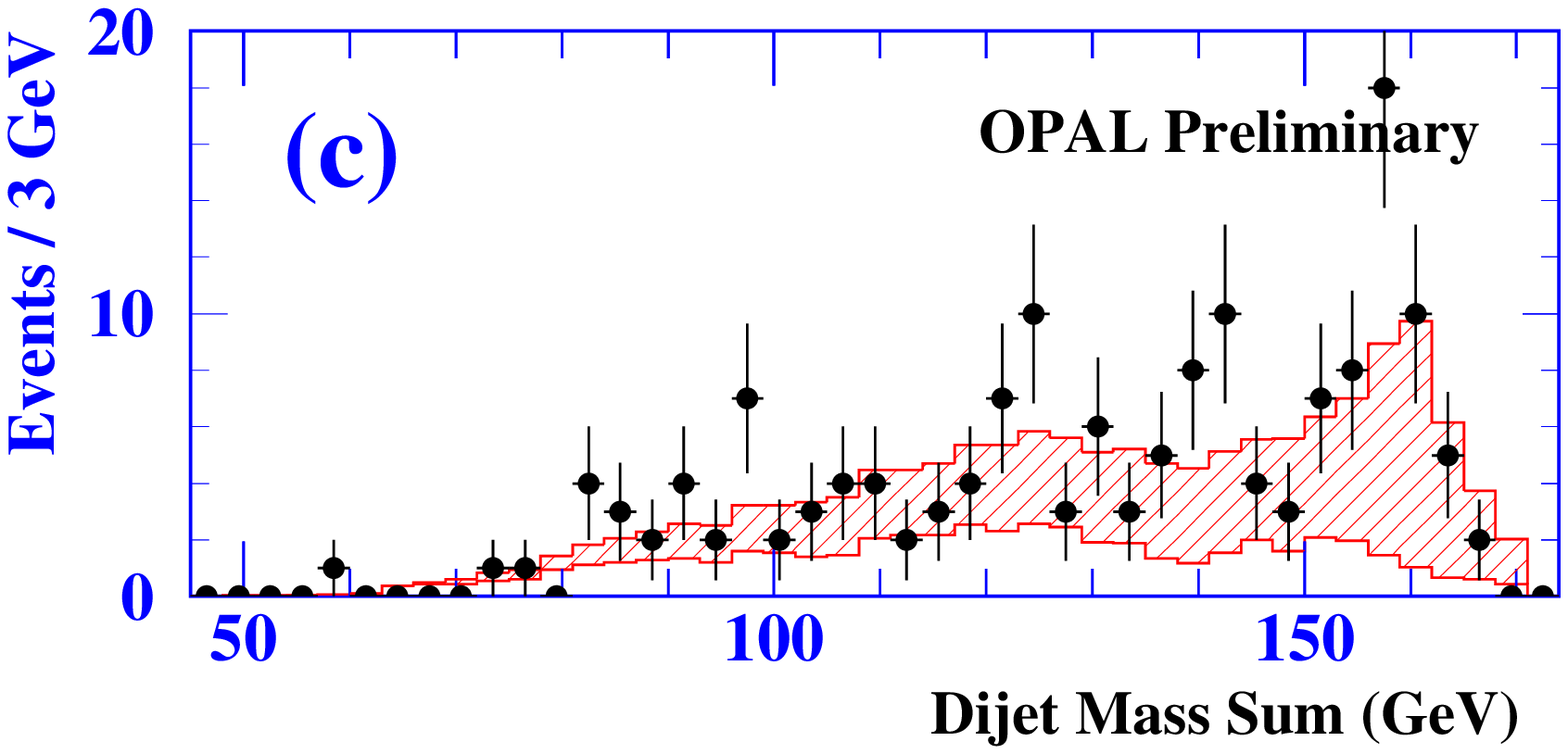,height=5cm,width=12.0cm}}
\caption{Distribution of the dijet mass sum in 4-jet events for OPAL data 
combined from all data samples ($\roots \geq 130$~GeV) after all cuts in the 
search for anomalous 4-jet production.  No
W-pair veto has been applied.  Plot~(a) shows the distribution for the 
combination with the minimum dijet mass difference,~$\Delta M$, plot~(b) 
accumulates the contributions from combinations for the smallest and the 
intermediate~$\Delta M$, and plot~(c) contains contributions from all three 
combinations.  Data are shown by points and SM backgrounds by the histogram.  
The hatched component of the background histogram denotes four-fermion 
processes (predominantly $\mrm{W^{+}W^{-}}$), while the unhatched component 
denotes $\mrm{Z}^{0}/\gamma \ra \qqbar$.}
\label{OPAL_4jet_plot}
\end{center}
\end{figure}

Using LEP1.5 data, the ALEPH collaboration reported a large excess of four-jet
events~\cite{aleph_4jets}. We have performed an analogous analysis sensitive 
to anomalous four-jet production.  Using a sample of simulated SUSY 
$\mrm{h^{0}A^{0}} \rightarrow \mrm{qqqq}$ ($\roots = 133$~GeV) as a benchmark 
for comparison, 
we achieve the same efficiency and background and a comparable mass 
resolution (to within $10\%$) as the ALEPH analysis, thus ensuring that the two
analyses have the same sensitivity.  For each event passing the cuts, the 
invariant mass of each jet-jet pair is calculated for all possible jet-jet
combinations.  The sum of the di-jet masses for that combination yielding the 
minimum mass difference between the two pairings is shown in 
Figure~\ref{OPAL_4jet_plot}, which includes all data taken at the 
center-of-mass
energies $133$, $161$, and $172$~GeV. We expect $26.0$ events and observe $20$.
The distribution of the sum of the di-jet masses is consistent with the 
SM background expectation. If systematic effects are neglected, the ALEPH and 
OPAL data are consistent at the level of $\sim 10^{-6}$. It should be noted 
that the inclusion of the systematic effects would reduce this significance.
 
\section{ {\bf $\mrm{WW}$ Physics at LEP2 }}
\label{ww}

At the center-of-mass energy $\roots = 161$~GeV the $\WW$ production 
cross-section is dominated by the so called ``CC03'' diagrams~\cite{ybook}: 
$s$-channel $\gamma$ or $\mrm{Z^{*}}$ exchange, and $t$-channel neutrino 
exchange. This center-of-mass energy lies just above the W pair production
threshold, and the cross-section here has a particularly strong dependence on 
the value of the mass of the W-boson, $\Mw$, so that it is possible to extract
$\Mw$ from the data by measuring the cross-section and comparing with 
theoretical predictions in the context of the SM.  These measurements are 
complementary to those at the Tevatron collider~\cite{tev_mw} and to those
which will be performed during the later phases of LEP2 operation by directly 
reconstructing the W decay products.  In addition, the two $s$-channel 
contributions to the cross-section are sensitive to the triple gauge 
couplings, $WWZ$ and $WW\gamma$.

\subsection{\bf Measurement of the W Boson Mass}
\label{mw_measurement}

The analysis is sensitive to all expected decay topologies, the fully hadronic
decays, $\WW \rightarrow \qqqq$, the semi-leptonic decays,
$\WW \rightarrow \qqln$, and the fully leptonic decays, $\WW \rightarrow \lnln$
($\ell$ = $e$, $\mu$, or $\tau$).  The dominant background is 
$\mrm{Z^{0}}/\gamma \rightarrow \mrm{f\overline{f}}$, where $\mrm{f}$ is any 
charged fermion.  Other backgrounds arise from four-fermion processes which do
not contain two resonant W bosons in the intermediate state.  These 
four-fermion backgrounds fall into two classes: those which can interfere with 
the $\WW$ four-fermion states, and those which cannot.  The interfering 
four-fermion backgrounds are particularly problematic because they can also 
depend on $\Mw$.  This mass-dependent four-fermion background is taken into 
account when extracting $\Mw$ from the observed data.  In addition, the 
cross-section for the process $\epem \rightarrow \WW$, arising from the CC03 
diagrams, is also measured from the data assuming that the interference terms
have only a small effect on the accepted $\WW$ cross-section.  This is a 
reasonable assumption given the current level of statistical precision.

Fully hadronic $\WW \rightarrow \qqqq$ events are selected as 
high-multiplicity, spherical, four-jet events, whose kinematics are compatible 
with the $\epem \rightarrow \WW$ hypothesis.  The semi-leptonic
$\WW \rightarrow \qqln$ events are characterized by two, high-multiplicity, 
back-to-back jets, an energetic lepton candidate (a low multiplicity jet in the
case of $\ell = \tau$), and large missing transverse momentum due to the 
escaping neutrino. The fully-leptonic decays, $\WW \rightarrow \lnln$ are 
selected as energetic, acoplanar, lepton pairs with large missing transverse 
momentum.  The efficiencies, and expected number of signal (assuming the world
average W-boson mass~\cite{wmass_ave}) and background events for each 
channel are given in Table~\ref{ww_table}.  Summing over all channels, we 
expect $27.6 \pm 2.5$ events and observe $28$.
\begin{table}
\begin{center}
  \begin{tabular}{|c|c|c|c|c|c|} \hline
            &           &\multicolumn{3}{c|}{expected} &    \\ \cline{3-5}
    Channel &Eff ($\%$) &Signal          &Bkgd &Total &Obs  \\ \hline \hline
    $\mrm{qqqq}$ 
      &$57$ &$9.6$ &$3.4$ &$13.0 \pm 1.1$ &$14$  \\
      & & & & & \\
    $\mrm{qqe}\nu$ 
      &$71$ &$3.9$ &$0.2$ &$4.1 \pm 0.5$ &$3$    \\
      & & & & & \\
    $\mrm{qq}\mu\nu$ 
      &$77$ &$4.2$ &$0.2$ &$4.5 \pm 0.5$ &$2$    \\
      & & & & & \\
    $\mrm{qq}\tau\nu$ 
      &$42$ &$2.3$ &$1.0$ &$3.2 \pm 0.4$ &$7$    \\
      & & & & & \\
    $\ell\nu\ell\nu$ 
      &$65$ &$2.6$ &$0.2$ &$2.8 \pm 0.3$ &$2$    \\
      \hline\hline
      & & & & & \\
    Total
      &$61$ &$22.6$ &$5.0$ &$27.6 \pm 2.5$ &$28$ \\
      \hline
  \end{tabular}
\end{center}
\label{ww_table}
\caption{ The efficiency, background, and number of observed events for each
  of the WW final state topologies.  The efficiency is calculated assuming the
  world average W mass and taking as signal only the CC03 diagrams.  The 
  backgrounds correspond to all other diagrams and assumes that the 
  interference effects can be neglected.}
\end{table}

By neglecting the $\Mw$ dependence of the interfering four-fermion backgrounds,
we can measure the W-pair (CC03) production cross-section using the 
information from each channel separately.  For each channel, the 
probability of obtaining the number of observed events is calculated as a 
function of the $\WW$ cross-section using Poisson statistics and assuming SM 
branching ratios.  A likelihood is formed from the product of the Poisson 
probabilities for each channel.  The maximum likelihood value yields a CC03 
cross-section of
\begin{equation}
  \sigma_{\mrm{WW}} = 
    3.62^{+0.93}_{-0.82}(\mrm{stat}) \pm 0.16(\mrm{syst})\mrm{pb}.
\end{equation}
The systematic uncertainty is evaluated by means of repeated MC trials.  The 
procedure takes into account the correlated luminosity uncertainties and the
small correlated systematic uncertainties between the semi-leptonic channels.

\begin{figure}[p]
\begin{center}
\begin{minipage}[t]{5in}
\epsfxsize = 5in
\epsffile{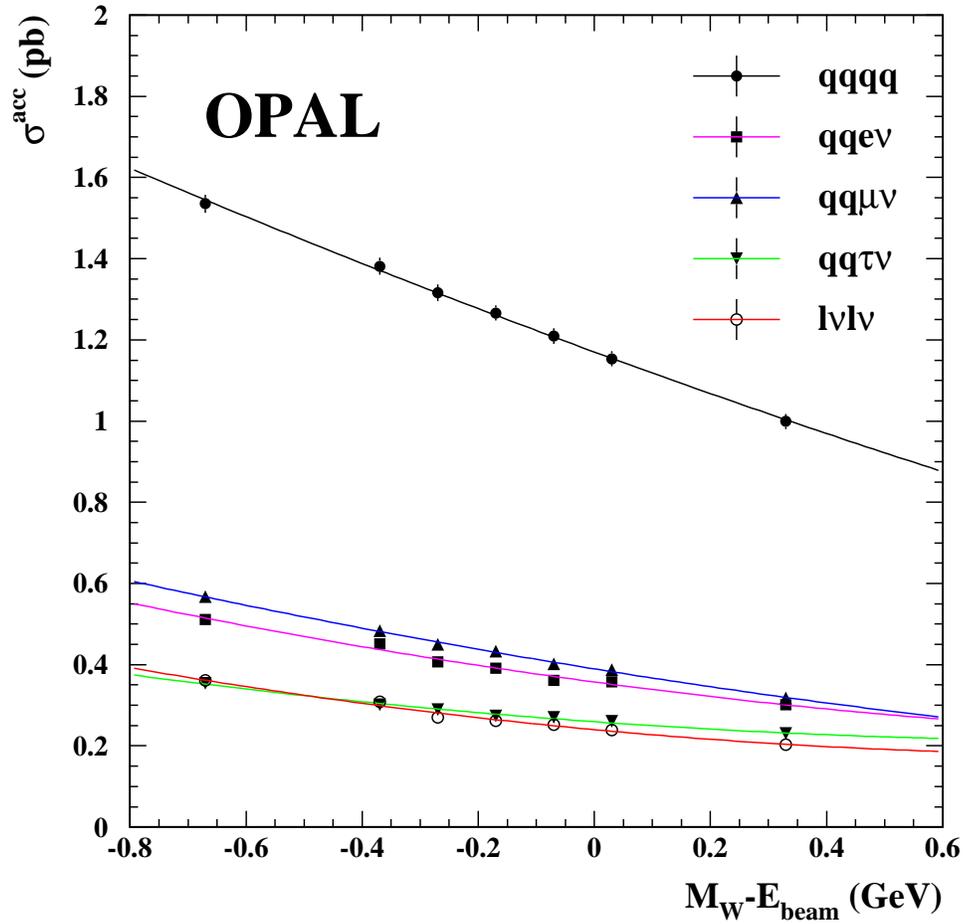}
\end{minipage}
\end{center}
\caption{The accepted cross-sections for each $\wpwm$ decay channel as a 
function of $\mrm{M_{W}} - \mrm{E_{beam}}$.  In each case the functional 
dependence is parametrised by a second order polynomial.  These are used in 
extracting the mass of the W boson and implicitly account for the 
four-fermion interference effects.}
\label{accxs_v_mw_plot}
\end{figure}

\begin{figure}[p]
\begin{center}
\begin{minipage}[t]{5in}
\epsfxsize = 5in
\epsffile{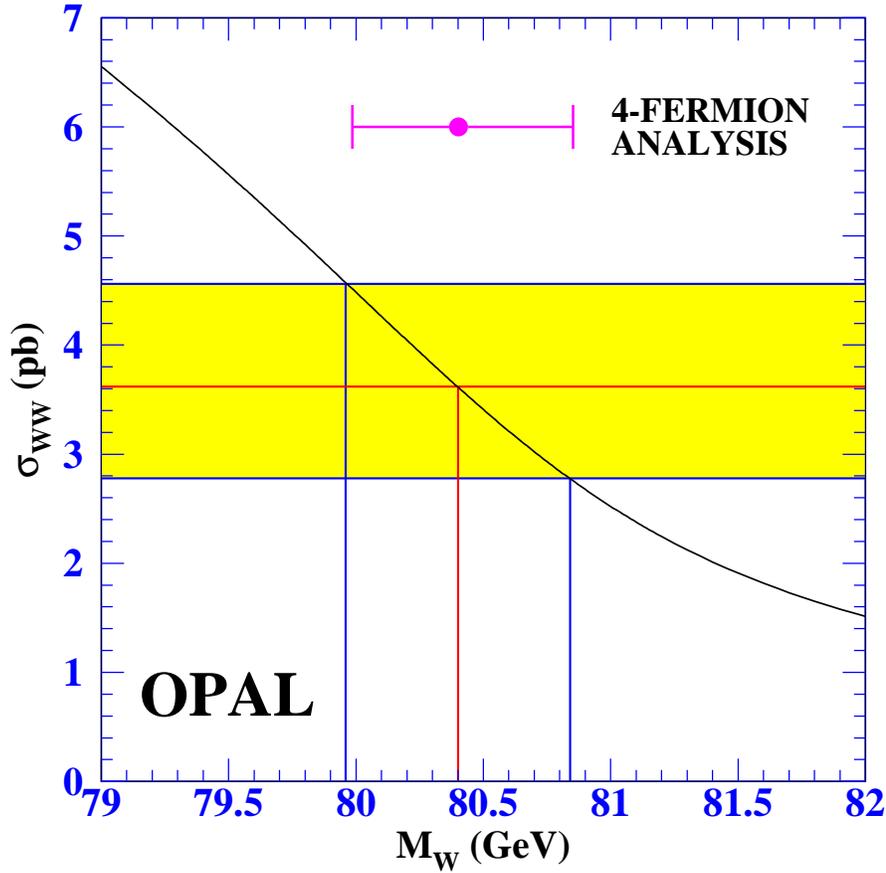}
\end{minipage}
\end{center}
\caption{Distributions of $\sigma_{WW}$ as a function of $\mrm{M_{W}}$ as 
predicted by the semi-analytic program GENTLE for $\roots = 161.3$~GeV.  The
measured $\mrm{W^{+}W^{-}}$ cross-section is shown as a shaded band and the 
corresponding W boson mass by vertical lines.  For comparison, our principal 
measurement of $\mrm{M_{W}}$ is shown as a point with error bars.  The 
uncertainties include statistical and systematic contributions, but do not
include the effect of the beam energy uncertainty.}
\label{cc03_plot}
\end{figure}

To determine the W-boson mass we parameterize the total accepted cross-section 
for each channel, including the effects of interfering four-fermion final 
states, as a function of $\Mw$ as shown in Figure~\ref{accxs_v_mw_plot}.  We 
employ a maximum likelihood technique analogous to the one described above to
determine
\begin{equation} 
  \Mw = 80.80^{+0.44+0.09}_{-0.41-0.10} \pm 0.03 \mrm{GeV},
\end{equation} 
where the first and second uncertainties are statistical and systematic,
respectively, and the third arises from the current estimate of the LEP beam
energy uncertainty.  As a cross-check, the  value of $\Mw$ can also be 
determined from the CC03 cross-section measurement described above by employing
the semi-analytic program GENTLE~\cite{gentle_ref} to derive the dependence of 
$\sigma_{\mrm{WW}}$ on $\Mw$, and by assuming that the experimental acceptance 
does not significantly vary as a function of $\Mw$.  The $\WW$ cross-section 
and resulting $\Mw$ measured in this CC03 framework are shown in 
Figure~\ref{cc03_plot}.  This measurement is consistent with the value 
determined in the full four-fermion analysis.

This analysis is described in more detail in Reference~\cite{mw_paper}.

\subsection{\bf Measurement of the Triple Gauge Couplings}
\label{tgc_measurement}

Anomalous triple gauge couplings (TGCs) can affect both the total production 
cross-section and the shape of the differential cross-section as a function of
the $\mrm{W}$ production angle.  The relative contributions of each helicity 
state of the W-bosons are also changed, which in turn affects the distributions
of their decay products.  

The most general Lorentz invariant Lagrangian has up to 14 independent 
$\mrm{WWV}$ couplings.  Requiring electromagnetic gauge invariance and 
$\mrm{C}$ and $\mrm{P}$ invariance reduces this parameter set to five, 3 
describing the $\mrm{WWZ}$ vertex and 2 descibing the $\mrm{WW}\gamma$ vertex.
This parameter space can be further reduced by considering constraints 
available from lower energy data and precise measurements at LEP1~\cite{ybook}.
As a result of these considerations, three specific linear combinations of 
these couplings have been proposed which are not tightly constrained by the 
lower energy data.  These are:
\begin{eqnarray*}
  \alpha_{B\phi} & \equiv 
                 & \Delta\kappa_{\gamma} - \Delta g^{z}_{1}\cos ^{2}\theta_{w} 
                   \\
  \alpha_{W\phi} & \equiv
                 & \Delta g^{z}_{1}\cos ^{2}\theta_{w} 
                   \\
  \alpha_{W}     & \equiv
                 & \lambda_{\gamma}
\end{eqnarray*}
with the constraints that 
$\Delta\kappa_{z} = \Delta g^{z}_{1} - \Delta \kappa_{\gamma}tan^{2}\theta_{w}$
where the $\Delta$ indicates the deviation of the respective quantity from the 
SM expectation and $\theta_{w}$ is the weak mixing angle.  We are most 
sensitive to the $\mrm{W}\phi$ model, which is the only model we presently 
consider assuming $\alpha_{B\phi}$ and $\alpha_{W}$ are zero.

We use both the total cross-section and relevant differential kinematic 
distributions to set limits on $\alpha_{W\phi}$.  For the cross-section 
analysis, the same selections are used as described in 
Section~\ref{mw_measurement}.  For the analysis of the kinematic distributions
we use only the $\qqln$ channels since - in contrast with the $\qqqq$ channel -
there is neither an ambiguity in assigning decay fermion pairs to each W, nor 
in determining the charges of each W.  These selections are augmented in order 
to further reduce the background.  The kinematic variables used are:
\begin{enumerate}
  \item $\cos\theta_{W}$, the production angle of the $\mrm{W^{-}}$ with 
    respect to the $\mrm{e^{-}}$ beam direction,
  \item $p_{W}$, the momentum of the hadronically decaying W
  \item $\cos\theta^{*}_{\ell}$, the polar decay angle of the charged lepton 
    with respect to the W flight direction measured in the W rest frame
  \item $\phi^{*}_{\ell}$, the azimuthal decay angle of the charged lepton with
    respect to a plane defined by the W and the beam axis.
\end{enumerate}

\begin{figure}[p]
\begin{center}
\begin{minipage}[t]{5in}
\epsfxsize = 5in
\epsffile{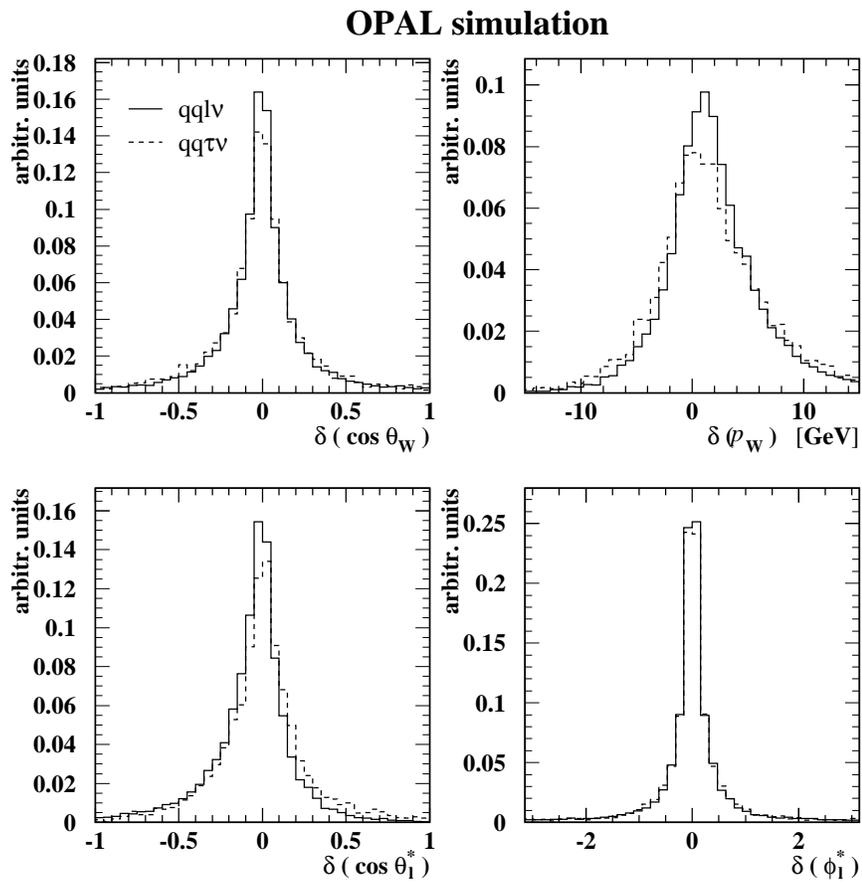}
\end{minipage}
\end{center}
\caption{The resolution of the kinematic variables from the $\qqln$ events used
in the TGC analysis.  All distributions show the difference between 
reconstructed and generated quantities.  The solid line is for $\qqen$ and 
$\qqmn$ events and the dashed line is for $\qqtn$ events. }
\label{tgc_qqln_kvar_plot}
\end{figure}

In the case of the $\qqen$ and $\qqmn$ channels we use variables resulting from
a kinematic fit demanding energy and momentum conservation.  For the $\qqtn$ 
channel we use energy and momentum constraints to calculate the energy of the 
of the $\tau$ where the $\tau$ flight direction is approximated by the 
direction of its observed decay products.  As demonstrated in 
Figure~\ref{tgc_qqln_kvar_plot}, the resolution of the kinematic variables, 
as estimated from MC, is comparable for all the $\qqln$ channels.

\begin{figure}[p]
\begin{center}
\begin{minipage}[t]{5in}
\epsfxsize = 5in
\epsffile{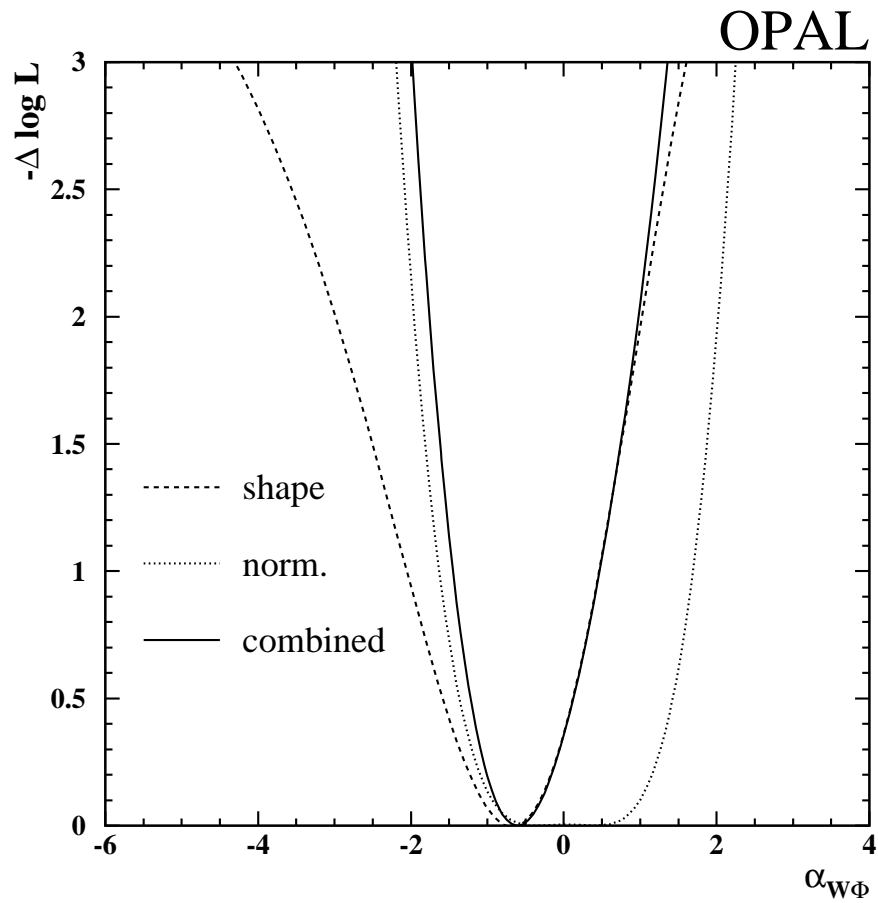}
\end{minipage}
\end{center}
\caption{Likelihood distributions obtained from the cross-section (dotted) 
and differential distributions (dashed) for the TGC analysis. The solid line is
the distribution obtained by adding these together.  In all cases the minimum 
value of the negative log likelihood has been subtracted.}
\label{tgc_lhood_plot}
\end{figure}

The total cross-section measurement is used to calculate a likelihood, 
analogous to the one described in Section~\ref{mw_measurement}, except that 
$\sigma_{WW}$ is parameterized as a function of $\alpha_{W\phi}$, assuming 
the world average $\Mw$.  For the differential distributions, we calculate the 
likelihood for the observed $\qqln$ events to have their measured distributions
of the kinematic variables as a function of $\alpha_{W\phi}$.  These 
likelihoods are independent and are added together to yield a total likelihood 
distribution, shown in Figure~\ref{tgc_lhood_plot}, from which we measure
\begin{equation}
  \alpha_{W\phi} = -0.61^{+0.73}_{-0.61} (\mrm{stat}) \pm 0.35 (\mrm{syst}).
\end{equation}
The corresponding $95\%$ confidence level limits are
\begin{equation}
  -2.1 < \alpha_{W\phi} < 1.6
\end{equation}
This analysis is described in more detail in Reference~\cite{tgc_paper}.

\section{\bf Summary}
\label{summary}

During the 1996 data taking run LEP ran for the first time at center-of-mass 
energies above the W-pair production threshold.  This new energy regime 
offers new tests of the SM, opens up a previously unexplored region of 
parameter space for a wide variety of models beyond the SM, such as SUSY, and 
affords the first study of $\WW$ events from which we can measure $\Mw$ and 
extract limits for anomalous triple gauge couplings.  OPAL has established a 
wide and varied physics program exploiting these opportunities 
\cite{qcd_paper} \cite{2ferm_paper} \cite{searches} \cite{sm_higgs_paper}
\cite{charg_neut_paper} \cite{mw_paper} \cite{tgc_paper}. 


\pagebreak

\pagebreak


\end{document}